\documentclass[prl,twocolumn,amsmath,amssymb]{revtex4}

\def\gsim{ \lower .75ex \hbox{$\sim$} \llap{\raise .27ex \hbox{$>$}} }

\def\lsim{ \lower .75ex \hbox{$\sim$} \llap{\raise .27ex \hbox{$<$}} }

\def\IZ{\relax\ifmmode\mathchoice
{\hbox{\cmss Z\kern-.4em Z}}{\hbox{\cmss Z\kern-.4em Z}}
{\lower.9pt\hbox{\cmsss Z\kern-.4em Z}} {\lower1.2pt\hbox{\cmsss
Z\kern-.4em Z}}\else{\cmss Z\kern-.4em Z}\fi}
\def\IR{\relax{\rm I\kern-.18em R}}

\def\one{{\hbox{ 1\kern-.8mm l}}}

\newlength{\bredde}
\def\slash#1{\settowidth{\bredde}{$#1$}\ifmmode\,\raisebox{.15ex}{/}
\hspace*{-\bredde} #1\else$\,\raisebox{.15ex}{/}\hspace*{-\bredde}
#1$\fi}

\newsavebox{\zzzbar}
\sbox{\zzzbar}
  {\setlength{\unitlength}{0.9em}
  \begin{picture}(0.6,0.7)
  \thinlines
  \put(0,0){\line(1,0){0.6}}
  \put(0,0.75){\line(1,0){0.575}}
  \multiput(0,0)(0.0125,0.025){30}{\rule{0.3pt}{0.3pt}}
  \multiput(0.2,0)(0.0125,0.025){30}{\rule{0.3pt}{0.3pt}}
  \put(0,0.75){\line(0,-1){0.15}}
  \put(0.015,0.75){\line(0,-1){0.1}}
  \put(0.03,0.75){\line(0,-1){0.075}}
  \put(0.045,0.75){\line(0,-1){0.05}}
  \put(0.05,0.75){\line(0,-1){0.025}}
  \put(0.6,0){\line(0,1){0.15}}
  \put(0.585,0){\line(0,1){0.1}}
  \put(0.57,0){\line(0,1){0.075}}
  \put(0.555,0){\line(0,1){0.05}}
  \put(0.55,0){\line(0,1){0.025}}
  \end{picture}}

\newcommand{\ena}{\end{eqnarray}}
\newcommand{\beqa}{\begin{eqnarray}}
\newcommand{\eeqa}{\end{eqnarray}}
\newcommand{\bea}{\begin{eqnarray}}
\newcommand{\eea}{\end{eqnarray}}

\newcommand{\be}{\begin{equation}}
\newcommand{\ee}{\end{equation}}

\usepackage{graphicx}

\def\ben{\begin{equation}}
\def\een{\end{equation}}

\def\bea{\begin{eqnarray}}
\def\eea{\end{eqnarray}}
\def\be{\begin{equation}}
\def\ee{\end{equation}}
\def\beq{\begin{eqnarray}}
\def\eeq{\end{eqnarray}}
\def\ba{\begin{eqnarray}}
\def\ea{\end{eqnarray}}

\input amssym.def
\input amssym.tex

\usepackage{dcolumn}

\begin{document}


\title{Universal properties and the first law of black hole inner mechanics}

\author{\bf Alejandra Castro$^{a}$} 
\email{acastro@physics.mcgill.ca}
\author{\bf Maria J. Rodriguez$^{b}$\,} 
\email{mjrodri@physics.harvard.edu}

\affiliation{
\vskip .5mm
{\it$^a$ Physics Department, McGill University, 3600 rue University, Montreal, QC H3A 2T8, Canada}  \vskip .5mm              
{\it $^b$ Center for the Fundamental Laws of Nature, Harvard University, Cambridge 02138, MA USA }
\vskip 0.5mm}

\begin{abstract} 
  We show by explicit computations that the product  of all the horizon areas is independent of the mass, regardless of the topology of the horizons. The  universal character of this relation holds for all known five dimensional asymptotically flat black rings, and  for black strings. This gives further evidence for the crucial role that the thermodynamic properties  at each horizon  play in understanding the entropy  at the  microscopic level. To this end we propose a ``first law" for the  inner Cauchy horizons of black holes. The validity of this formula, which seems to be universal, was explicitly checked in all cases.
\end{abstract}

\maketitle 
 
 \section{Introduction}
The establishment  that laws of black holes mechanics are thermodynamical laws \cite{Bardeen:1973gs,Hawking:1974sw}  has raised  challenging questions.  A statistical derivation of the thermal properties of the black hole, and hence an accounting for its microstates,  remains puzzling.  Within string theory, the seminal work of \cite{Strominger:1996sh} identified precisely the microscopic degrees of freedom for a supersymmetric -- zero temperature -- black hole. Despite the narrowness of the construction, it provided a framework to explore the robustness of the result beyond its initial scope. Since then we have outgrown our insights to move the discussion towards a larger class of black holes, including solutions at finite temperature \cite{Guica:2008mu,CastroMaloneyStrominger}. The core idea of this program is not in string theory, but rather guided by the holographic principle: there is a non-trivial match between features of a 2d conformal field theory (CFT$_2$) and features of black holes.

Part of our current understanding of  black hole thermodynamics relies essentially on the universal symmetry arguments that follow from the analysis of  3d gravity in AdS spacetime \cite{Brown:1986nw,Strominger:1997eq}. The microscopic degrees of freedom of the black hole are described in terms of those of a conformal field theory -- without gravity-- living in the boundary. A by-product of the construction is that the area product of the inner $A_-$ and outer $A_+$ Killing horizons of black hole  gives 
\ben \label{algo2}
 {A_+\,A_-\over (8\pi G_3)^2} =N_R-N_L~,
\een
where $N_{R}, N_{L}$ are respectively the number of right and left moving excitations of the CFT$_2$. The left hand side of this equation is the level matching condition of the CFT, i.e. the requirement that the momentum along a compact spatial direction is quantized.      

Interestingly, it seems to be the case that {\it any} asymptotically flat black hole in $d$-spacetime dimensions, which admits a smooth extremal limit, satisfies \cite{Larsen:1997ge}
\ben \label{algo}
{A_+\,A_-\over (8\pi G_d)^2} \in \mathbb{Z} ~,
\een
resembling  \eqref{algo2}. The precise statement is that the products of areas is independent of the mass of the black hole and therefore  depends solely on the quantized charges. This suggestive general relation for black holes prompted the purely gravitational investigation of the product of areas of black holes \cite{Cvetic:2010mn}, where the topology of the horizon is a sphere. 
The product of areas had also been investigated for  Kerr-Newman black holes in \cite{Ansorg:2008bv,Ansorg:2009yi}.

Higher dimensional gravity allows for  more exotic species of regular solutions. It is natural then to enquire whether the simple relation \eqref{algo} is truly universal and hence satisfied for less symmetric non-spherical black holes such as black rings and black strings. By direct computation for all known non-spherical asymptotically flat black holes in five dimensions we find that the relation is indeed independent of the mass.
We take this observation as a starting point to construct the microscopic structure for  black rings and black strings, which we will elaborate below.  

Contrary to the outer-horizon thermodynamics, it is not clear whether 
the inner-horizon has any relevance for a statistical accounting of the black hole entropy; our first hint is \eqref{algo}.
However, as we show, one can similarly consider the inner thermodynamics of black holes.
In particular, there is a ``first law" for the inner-horizon that is valid for all the black 
solutions that we have so far considered. The first law of the black hole inner mechanics is schematically
\be\label{first}
-\,dM=T_-\frac{d A_-}{4G_5} \,-({\Omega^-} \,d J + \Phi_{E}^{-}\,dQ +  \Phi^{-}_{m}\, dq)~,
\ee
where the extensive quantities are the ADM charges and the corresponding intensive quantities
are defined at the inner Cauchy horizon.  Ref.  \cite{Curir1,Curir2} studied this relation for the Kerr black hole and for more general black holes in \cite{Cvetic1,Cvetic2}; we show that (\ref{first}) holds regardless of the complexity of the solution.  For the sake of comparison, the first law for the outer horizon is
\be\label{firstevent}
dM=T_+\frac{d A_+}{4G_5} \,+({\Omega^+} \,d J + \Phi_{E}^{+}\,dQ +  \Phi^{+}_{m}\, dq)\,.
\ee
Contrasting equations (\ref{first}) and (\ref{firstevent}), we observe that up to signs they are essentially the same after exchanging the definitions of the potentials and areas at the inner and outer horizons.  The minus signs  in (\ref{first}) are due to the Killing horizon vector field being space-like inside the black hole event horizon.  We are assigning negative energy ($-M$) to the inner horizon, similar to the negative energies within the ergosphere.  %

Classically the inner horizon is perturbatively unstable; the physical implications of this instability have been recently revisited in \cite{Marolf:2011dj}. 
For our purposes, the inner horizon is a mathematical artifact that provides an interesting venue. Our aim is to revive the topic and illustrate how it could impact a statistical interpretation of black hole thermodynamics.

\section{Definitions}
\label{sec:Def}

Our focus is on thermodynamics properties of solutions which have smooth horizons with non-spherical topologies.
We require that the solutions have a regular inner $(r_-)$ and outer $(r_+)$ horizon, and admit a smooth extremal limit $r_+=r_-$ with vanishing Hawking temperature. Known analytic solutions in 5d supergravity are the neutral and charged rotating black rings   \cite{Pomeransky:2006bd,Emparan:2004wy,Elvang:2004xi}; the horizons have topology $S^1\times S^2$. 

To begin we first define the quantities appearing in \eqref{first}.  The extensive variables $(M,J_\phi,J_\psi,Q_i,q_i)$ are the ADM charges defined at asymptotic infinity. Strictly speaking the dipole charge $q_i$ is not a conserved charge; its role in the thermodynamics of the black ring was discussed in detail in  \cite{Copsey:2005se}. These are the same definitions as in \cite{Elvang:2004xi}, with the exemption that charges are measured in Planck units 
\be
Q_i\to \ell_P^2 Q_i~, \quad q_i\to \ell_P q_i~,  \quad \ell_P^3 \equiv{\pi\over4 G_5}~.
\ee 
The index $i=1,2,3$ labels the $U(1)^3$ charge that the supergravity solution carries.

The intensive variables are intrinsic to each horizon. In all cases the subscripts indicate the values of the intensive quantities and areas at the outer $(+)$ or inner $(-)$ horizons. 
To compute them, the ADM  form is convenient where we have 
\ben
ds^2= -N^2dt^2 + \gamma_{ab}(dx^a + N^adt )(dx^b + N^b dt )~,
\een
with $x^a$ four spatial directions; $N(x^a)$ and $N^b(x^a)$ are the lapse function and the shift vector respectively.  The horizons are defined as the zeroes of the appropriate radial component of the metric.  
As in \cite{Astefanesei:2010bm}, the angular potentials and temperatures for each horizon are defined 
\ben\label{genpot}
\Omega_{k}^{\pm} =- \left. N^{k}\right|_{r_\pm}   ~,\quad T_{\pm} ={1\over 4\pi} \left| {(N^2)' \over \sqrt{g_{rr}N^2}}\right|_{r_{\pm}}~,
\een
where $k=1,2,...,[\frac{d-1}{2}]$.
With this definition $T_+$ corresponds to the Hawking temperature. A semiclassical physical interpretation of  $T_-$ is subtle. In analogy, however, we can define a geometrical positive quantity suitable for (\ref{first}) that we will call the ``inner temperature" and is constant over the inner horizon.

The solutions under consideration have three Killing vectors: a time-like vector $\partial_t$, $\partial_\phi$ and $ \partial_\psi$.   The Killing vectors that define the inner and outer horizon are then
\ben
\chi_{\pm}=\partial_t -\Omega_{\psi}^{\pm}\,\partial_{\psi}-\Omega_{\phi}^{\pm}\,\partial_{\phi}~,
\een
where the coordinates $({\psi},{\phi})$ as those with periodicity $2\pi$. 

The solutions  are supported as well  by  one-form potentials $A_\mu^i$, which due to the non-spherical shape of the horizon create both electric and dipole moments.  The electric potential for each horizon is
\ben
\label{elecpot}
\Phi_{E,i}^{\pm}=\ell_P \left[(\chi^\mu A_\mu^i)_\infty -  (\chi^\mu A_\mu^i)_{ r_{\pm}}\right]~.
\een
To compute the dipole potential we employ the formula presented in \cite{Copsey:2005se}. In the absence of electric charges, this reduces  to
\ben
\Phi_{m,i}^{\pm}= \ell_P^{2}[(A_{\phi}^i)_{\infty}-(A_{\phi}^i)_{r_{\pm}}]~.
\een

\section{Universality of $A_+A_-$}
Our first observation is the universal behavior of  product of the horizon areas for black rings in $d=5$.   Currently there are 3 classes of known black ring solutions -- that are not smoothly connected to each other -- and satisfy the requirements specified in the previous section.    

First, consider the  neutral doubly spinning black ring \cite{Pomeransky:2006bd} with mass $M$, and two angular momenta: $J_\psi$ along  $S^1$ and $J_\phi$  on $S^2$. Evaluating  \eqref{algo} gives
\be {A_{+}A_{-}\over (8\pi G_5)^2}= J_\phi^2~.\ee
The product is not only independent of $M$, but also $J_\psi$. 

The second class is electrically and magnetically charged black rings \cite{Emparan:2004wy,Elvang:2004xi}. The most general solution carries electric $(Q_i)$ and dipole $(q_i)$ charge in addition to rotation $(J_\psi)$ along $S^1$. We find
\bea\label{AAST} &&{A_{+}A_{-}\over (8\pi G_5)^2}=J_\psi q_1q_2q_3
-{1\over 4}[(Q_1q_1)^2+(Q_2q_2)^2+(Q_3q_3)^2] \cr &&+{1\over2}[Q_1q_1Q_2q_2+Q_2q_2Q_3q_3+Q_3q_3Q_1q_1] ~.\eea
A peculiarity  is that the solution is as well rotating along $S^2$, but it is not an independent parameter. 
Taking $Q_i=0$ and $q\equiv q_i$, the system reduces to the dipole black ring \cite{Emparan:2004wy}, and formula (\ref{AAST}) still holds in this limit. 

And the third class,  are  electrically and magnetically charged black string \cite{Compere:2010fm}. These are solution of minimal supergravity where $Q\equiv Q_i$ and $q\equiv q_i$ and there are two independent rotation parameters. We obtain
\bea
&&{A_{+}A_{-}\over (8\pi G_5)^2}=J_\psi q^3
+{3\over 4} (Qq)^2+J_{\phi}^2~.
\eea
See also \cite{Meessen:2012su} for further examples.
The solution is not strictly speaking a ring; still it should be the limiting solution of the most general black ring solution of minimal supergravity when its size $R\rightarrow \infty$. 

The most general black ring solution of $U(1)^3$ supergravity should be characterized by nine independent parameters  $(M,J_\phi,J_\psi,Q_i,q_i)$; unfortunately it has not been constructed. A five parameter solution is being scrutinized \cite{JMeO}. We expect no departure from (\ref{algo}) for other cases, and anticipate for the general case
\bea
{A_{+}A_{-}\over (8\pi G_5)^2}&=&J_{\phi}^2+J_\psi \prod _{i=1}^{3}q_i\\
&&-{1\over 4}\sum_{i=1}^{3}\,[(Q_iq_i)^2-Q_iq_i\sum_{j\ne i=1}^{3} Q_jq_j]\,,\nonumber
\eea
in agreement with the near extremal analysis of \cite{Larsen:2005qr}.

\section{First Law of black holes Inner mechanics}

Pursing our curiosity about the inner Cauchy horizon, and  the definitions in Section \ref{sec:Def}, we propose a first law for the mechanics of the inner horizon
\be\label{firstdetail}
-\,dM=T_-\frac{d A_-}{4\,G} -({\Omega_k^-} \,d J_k + \Phi_{E,i}^{-}\,dQ_i +  \Phi^{-}_{m,i}\, dq_i)\,,
\ee
that is satisfied for all known 5d black rings \cite{Pomeransky:2006bd,Emparan:2004wy,Elvang:2004xi}, black strings \cite{Compere:2010fm} and d-dimensional black holes \cite{Myers:1986un} (as well as AdS-black holes \cite{Gibbons:2004uw}).

To illustrate our computations we present three examples. The inner horizon area and extensive thermodynamical quantities at the inner horizon are new.
 \begin{figure}
  \includegraphics[height=40mm,width=65mm]{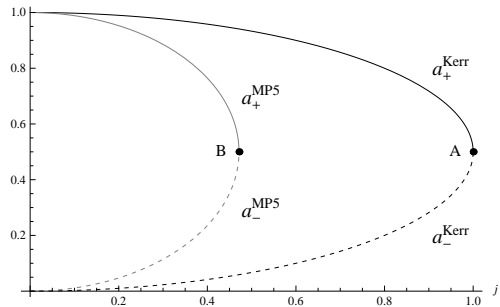}
    \caption{Phase diagram of black holes: $a_\pm$ vs. $j$ where $a_\pm=A_\pm/(16\pi M)$ and $j=J/M$. The {\it solid} curve shows the horizon area of the Kerr and  $5d$ Myers-Perry (with $J_1=J_2=J$) black hole while the {\it dashed}  line outlines the inner horizon area both as functions of angular momenta $j$ and fixed mass $M$. As $j$  increases the $a_-$ grows while $a_+$ decrease coinciding at the extremal value (points A, B).
}
\end{figure}

{\bf Black holes:} For AdS black holes \cite{Gibbons:2004uw} and Myers-Perry black holes \cite{Myers:1986un} ($L\rightarrow \infty$ below) in $d$-dim we find
\bea\label{ex1}
&&A_-=\frac{\Omega_{d-2}}{r_{-}^{1-\epsilon}}\prod_k \frac{(r_-^2+a_k^2)}{\Xi_k}\,,\,\, \Omega_k^-=\frac{a_k }{(r_-^2+a_k^2)\,\Lambda}\,,\nonumber\\
&&2\pi\, T_{-}=\frac{1}{r_-}-\frac{r_-}{\Lambda}\left(\sum_k\frac{1}{r_-^2+a_k^2}+\frac{\epsilon}{2 \,r_-^2}\right)\,,
\eea
where $\epsilon=(d-1)\mod 2$, $r_-$ is the smallest (positive) root of $r_-^{\epsilon-2}(1+r_-^2/L^2)\prod_k(r_-^2+a_k^2)=2m$,  $\Xi_k=1+a_k^2/L^2$  and  $\Lambda=(1+r_-^2/L^2)^{-1}$. We elaborate on the inner black hole mechanics in Fig. 1. The areas (for fixed mass) as functions of the angular momenta have mirroring behaviors.

{\bf Doubly spinning black ring}  \cite{Pomeransky:2006bd} : The solutions is parametrized by a scale $R$ and parameters $\nu$ and $ \lambda$ that satisfy $0\le\nu<1$ and $2\,\sqrt{\nu}\le \lambda<1+\nu$. The inner horizon is at $r_-=(-\lambda-\sqrt{\lambda^2-4\nu})/ 2\nu$.  Our results are
\bea\label{ex2}
&&A_{-}={32\pi^2 R^3}{(1+\lambda+\nu)\lambda\over \left(r_-^{-1}-r_-\right) (1-\nu)^2}\,,\nonumber\\
&&T_{-}=\frac{\left(r_-^{-1}-r_-\right)(1-\nu )\sqrt{\lambda ^2-4 \nu }}{8\pi R \lambda (1+\nu +\lambda )}\,,\\
&&\Omega_\psi^-=\frac{1}{2 R}\sqrt{\frac{1+\nu-\lambda}{ 1 +\nu+\lambda }}\,,\nonumber\\
&&\Omega_\phi^-=\frac{\lambda  (1+\nu )+(1-\nu )\sqrt{\lambda ^2-4 \nu }  }{4 R \lambda  \sqrt{\nu } }\sqrt{\frac{1+\nu-\lambda}{ 1 +\nu+\lambda }}\,.\nonumber
\eea

{\bf Dipole black rings:}  The black ring solution \cite{Emparan:2004wy} is parametrized by a scale $R$, the dipole parameter $\mu$ and $\lambda, \nu$ within $0<\nu\le\lambda<1$ and $0\le \mu<1$.
We find
\bea\label{ex3}
&&A_{-}=8\pi^2R^3{(1+\mu)^3 \over (1-\nu)^2}\sqrt{\mu^{3}\lambda(\lambda-\nu)(1-\lambda^2)}\,,\nonumber\\
&&\Omega_{\psi}^-=\frac{(1-\nu)}{R}\sqrt{\frac{\lambda}{(1+\mu)^{3}(1+\lambda)(\lambda-\nu)}}\,,\\
&&T_{-}=\frac{\nu}{4\pi R }\sqrt{1-\lambda\over \mu^3 (1+\lambda)(\lambda-\nu)}\,,\nonumber\\
&&\Phi^-_m={3R}\,\ell_P^2 {1+\mu\over 1-\nu}\sqrt{\frac{(\mu+\nu)(1-\mu)(1-\lambda)}{\mu}}\nonumber
\eea

With these results (\ref{ex1})-(\ref{ex3}) and respectively the ADM charges found in \cite{Gibbons:2004ai,Myers:1986un,Emparan:2004wy,Pomeransky:2006bd} we verified (\ref{firstdetail}).

We can also construct a Smarr relation for the inner horizon
\ben\label{smarr}
-\,M=\frac{d-2}{d-3}\left[T_- \,\frac{A_-}{4 G_d}-{\Omega^-} J\right] - \Phi^{-}_{E}\, Q-\frac{1}{2} \Phi^{-}_{m}\, q\,,
\een
which mimics  the Smarr relation for the outer horizon. It is straightforward to check this relation for all the asymptotically flat (AF) solutions of interest.  

\section{Conclusions}
\label{conclusions}
In this work, we have explored formulae for the products of the horizon areas  
for solutions with non-spherical horizon topology and consistently find that it is independent of the mass. We also observe that the inner Cauchy horizon of all solutions discussed above satisfies  the first law \eqref{firstdetail} and, for AF solutions, the Smarr relation \eqref{smarr}. 

 Besides this mathematical curiosity, from these relations  we can speculate on how to account microscopically for the entropy of black rings. There is accumulating evidence  that the Bekenstein-Hawking area law for spherical black holes agrees exactly with the microscopic degeneracy inferred from a CFT$_2$, i.e.
\be\label{cardy}
{A_+\over 4 G_d}= 2\pi \sqrt{N_R} +2\pi \sqrt{N_L}~.
\ee
 It is also the case that the inner horizon area satisfies ${A_-\over 4 G_d}= 2\pi \sqrt{N_R} -2\pi \sqrt{N_L}$.
The consistency and cohesiveness of this statistical derivation of the entropy is due in part to two facts. First, that $(A_+A_-)$ obeys \eqref{algo} as predicted by the CFT$_2$ description. Second, the first law \eqref{first} is consistent with the treatment of $N_{R,L}$ as weakly interacting sectors in the thermodynamic limit.\\  
 \indent
Extending this proposal to black rings has been challenging. Several of the ingredients used to obtain \eqref{cardy} in e.g. \cite{Larsen, CastroMaloneyStrominger}, are based on features that black holes do not share with their ring cousins.  To date, only in the (near-) extremal limit there is a sense in which we can still use a CFT$_2$ \cite{Larsen:2005qr,Reall:2007jv,Emparan:2008qn}. \\
\indent Nevertheless we can argue the following. Because all black rings admit two independent thermodynamical relations on each Cauchy horizon, nothing prevents us from writting ${A_{\pm}\over 4G_d}=S_{R}\pm S_L$. Hence we can pretend that the entropy comes from counting ``right'' and ``left'' movers with degeneracy $S_{R,L}$ respectively. And by construction, in the extremal limit where $A_+=A_-$ we have $S_L=0$. We can also make a well-educated guess and declare that $S_{L,R}=2\pi \sqrt{N_{L,R}}$.  This would naturally fit with \eqref{algo} and the results mentioned above for the extremal limit. To justify such a bold guess, we need to provide a gravitational definition of $N_{L,R}$; this can be done and will be carefully discussed elsewhere \cite{AAJMe}. 

In summary the two universal properties \eqref{algo} and \eqref{first} are necessary --but far from sufficient-- ingredients needed to account for the entropy via the statistical properties of a CFT$_2$. 
It would be interesting to test the applicability of our findings to more general theories and investigate the thermodynamic nature of the inner horizons.
 A natural extension is to de Sitter black holes and to explore the effects of the cosmological horizon in the construction presented here. 

\noindent
{\bf Acknowledgements}
We are grateful to D. Anninos, R. Emparan, G. Gibbons, G. Horowitz, T. Jacobson, J. Lapan, F. Larsen, A. Maloney, D. Marolf and A. Strominger
for useful discussions.  In addition we would like to thank the  Centro de Ciencias de Benasque Pedro Pascual, Spain for hospitality during the initial stages of this project. 
A. C.  research was  supported in part by NSERC and the NSF under Grant No. PHY11-25915.  M. J. R. was supported by PIOF-GA 2010-275082.

\nopagebreak


\begin{thebibliography}{42} 

\bibitem{Bardeen:1973gs} 
  J.~M.~Bardeen, B.~Carter and S.~W.~Hawking,
  Commun.\ Math.\ Phys.\  {\bf 31}, 161 (1973).

\bibitem{Hawking:1974sw} 
  S.~W.~Hawking,
  Commun.\ Math.\ Phys.\  {\bf 43}, 199 (1975)

\bibitem{Strominger:1996sh} 
  A.~Strominger and C.~Vafa,
  Phys.\ Lett.\ B {\bf 379}, 99 (1996)
  [hep-th/9601029].

\bibitem{Guica:2008mu} 
  M.~Guica, T.~Hartman, W.~Song and A.~Strominger,
  Phys.\ Rev.\ D {\bf 80}, 124008 (2009)
  [arXiv:0809.4266 [hep-th]].


\bibitem{CastroMaloneyStrominger}
A. Castro, A. Maloney and A. Strominger,
Phys. Rev. {\bf D82}, 024008 (2010), arXiv:1004.0996 [hep-th].

\bibitem{Brown:1986nw} 
  J.~D.~Brown and M.~Henneaux,
  Commun.\ Math.\ Phys.\  {\bf 104}, 207 (1986).

\bibitem{Strominger:1997eq} 
  A.~Strominger,
  JHEP {\bf 9802}, 009 (1998)
  [hep-th/9712251].
  
\bibitem{Larsen:1997ge}
 F.~Larsen,
 Phys.\ Rev.\  {\bf D56 } (1997)  1005-1008.
 [hep-th/9702153].

\bibitem{Cvetic:2010mn}
  M.~Cvetic, G.~W.~Gibbons, C.~N.~Pope,
  Phys.\ Rev.\ Lett.\  {\bf 106 } (2011)  121301.
  [arXiv:1011.0008 [hep-th]].

\bibitem{Ansorg:2008bv} 
  M.~Ansorg and J.~Hennig,
  Class.\ Quant.\ Grav.\  {\bf 25}, 222001 (2008)
  [arXiv:0810.3998 [gr-qc]].
  
\bibitem{Ansorg:2009yi} 
  M.~Ansorg and J.~Hennig,
  Phys.\ Rev.\ Lett.\  {\bf 102}, 221102 (2009)
  [arXiv:0903.5405 [gr-qc]].
  
\bibitem{Curir1} 
A.~Curir,
Nuovo Cimento B,{\bf 51B},  262 (1979).

\bibitem{Curir2} 
A.~Curir and M.~ Francaviglia,
Nuovo Cimento B,{\bf 52B},  165 (1979).

\bibitem{Marolf:2011dj} 
  D.~Marolf and A.~Ori,
  arXiv:1109.5139 [gr-qc].

\bibitem{Pomeransky:2006bd}
A.~A.~Pomeransky and R.~A.~Sen'kov,
 arXiv:hep-th/0612005.


\bibitem{Emparan:2004wy}
R.~Emparan,
JHEP {\bf 0403}, 064 (2004).
[hep-th/0402149].
  
\bibitem{Elvang:2004xi}
H.~Elvang, R.~Emparan and P.~Figueras,
 JHEP {\bf 0502}, 031 (2005)
  [hep-th/0412130].
  
\bibitem{Copsey:2005se} 
  K.~Copsey and G.~T.~Horowitz,
  Phys.\ Rev.\ D {\bf 73}, 024015 (2006)
  [hep-th/0505278].
  
\bibitem{Astefanesei:2010bm} 
  D.~Astefanesei, M.~J.~Rodriguez and S.~Theisen,
  JHEP {\bf 1008}, 046 (2010)
  [arXiv:1003.2421 [hep-th]].

\bibitem{Compere:2010fm} 
  G.~Compere, S.~de Buyl, S.~Stotyn and A.~Virumani,
  JHEP {\bf 1011}, 133 (2010)
  [arXiv:1006.5464 [hep-th]].

\bibitem{Meessen:2012su} 
  P.~Meessen, T.~Ortin, J.~Perz and C.~S.~Shahbazi,
  arXiv:1204.0507 [hep-th].

\bibitem{JMeO}
  J.~V.~Rocha, M.~J.~Rodriguez and O.~Varela,
  arXiv:1205.0527 [hep-th].
  

\bibitem{Myers:1986un} 
  R.~C.~Myers and M.~J.~Perry,
  Annals Phys.\  {\bf 172}, 304 (1986).
 
\bibitem{Gibbons:2004uw} 
  G.~W.~Gibbons, H.~Lu, D.~N.~Page and C.~N.~Pope,
  J.\ Geom.\ Phys.\  {\bf 53}, 49 (2005)
  [hep-th/0404008].
  
\bibitem{Gibbons:2004ai} 
  G.~W.~Gibbons, M.~J.~Perry and C.~N.~Pope,
  Class.\ Quant.\ Grav.\  {\bf 22}, 1503 (2005)
  [hep-th/0408217].
  
  \bibitem{Cvetic1}
  M.~Cvetic and F.~Larsen,
 Phys.\ Rev.\ D {\bf 56}, 4994 (1997)
 [hep-th/9705192].
 
\bibitem{Cvetic2}
M.~Cvetic and F.~Larsen,
 Nucl.\ Phys.\ B {\bf 506}, 107 (1997)
 [hep-th/9706071].
  
\bibitem{Larsen}
  F. Larsen,
Phys. Rev. {\bf D56} (1997) 1005, hep-th/9702153.
   
\bibitem{Larsen:2005qr}
  F.~Larsen,
  JHEP {\bf 0510 } (2005)  100.
  [hep-th/0505152].
  
\bibitem{Reall:2007jv} 
  H.~S.~Reall,
  JHEP {\bf 0805}, 013 (2008)
  [arXiv:0712.3226 [hep-th]].
  
\bibitem{Emparan:2008qn} 
  R.~Emparan,
  Class.\ Quant.\ Grav.\  {\bf 25}, 175005 (2008)
  [arXiv:0803.1801 [hep-th]].
  
  \bibitem{AAJMe}
 Work in progress, A.~Castro, J.~Lapan, A.~Maloney, M.~J.~Rodriguez. 
\end{thebibliography}
\end{document}